\documentclass[journal]{new-aiaa}
\usepackage[utf8]{inputenc}
\usepackage{textcomp}
\usepackage{multirow}
\usepackage{graphicx}
\usepackage{amsmath}
\usepackage{float}
\usepackage[version=4]{mhchem}
\usepackage{siunitx}
\usepackage{longtable,tabularx}
\usepackage{tikz-dimline}

\usepackage{xcolor}
\usepackage{placeins}
\definecolor{darkred}{HTML}{8B0000}   % --> darkred

\usepackage{soul}
\usepackage{cancel}

\usepackage{pgfplots}
\DeclareUnicodeCharacter{2212}{−}
\usepgfplotslibrary{groupplots,dateplot,external}
\usetikzlibrary{patterns,shapes.arrows}
\usetikzlibrary{shapes.geometric}
\pgfplotsset{compat=newest}
\usetikzlibrary{decorations}
\usetikzlibrary{decorations.markings}
\usetikzlibrary{positioning,calc}
\pgfplotsset{axis line style = thick} %<- you may want to add this

\newcommand\Rey{\mbox{\textit{Re}}}  % Stanton number

\makeatletter
\tikzset{
	nomorepostactions/.code={\let\tikz@postactions=\pgfutil@empty},
	mymark/.style 2 args={decoration={markings,
			mark= between positions 0 and 1 step (1/11)*\pgfdecoratedpathlength with{%
				\tikzset{#2,every mark}\tikz@options
				\pgfuseplotmark{#1}%
			},  
		},
		postaction={decorate},
		/pgfplots/legend image post style={
			mark=#1,mark options={#2},every path/.append style={nomorepostactions}
		},
	},
	closemark/.style 2 args={decoration={markings,
			mark= between positions 0 and 1 step (1/101)*\pgfdecoratedpathlength with{%
				\tikzset{#2,every mark}\tikz@options
				\pgfuseplotmark{#1}%
			},  
		},
		postaction={decorate},
		/pgfplots/legend image post style={
			mark=#1,mark options={#2},every path/.append style={nomorepostactions}
		},
	},
}
\makeatother

\title{Physics-Informed Acoustic Liner Optimization: \\Balancing Drag and Noise}

\author{Haris Shahzad, Stefan Hickel and Davide Modesti}
\affil{Aerodynamics Group, Faculty of Aerospace Engineering, Delft University of Technology,\\ Kluyverweg 2, 2629 HS Delft, The Netherlands}

\begin{document}
	
	\maketitle
	
	\begin{abstract}
		We present pore-resolved Direct Numerical Simulations (DNS) of turbulent flows
		grazing over acoustic liners with aerodynamically and/or acoustically optimized orifice configurations.
		Our DNS explore a large parameter space, studying various families of orifice geometries, including the influence of orifice shape, orientation,
		and the number of orifices.
		All flow cases show an increase in drag compared to the smooth wall.
		However, the added drag can be reduced by as much as $\sim$55\% as compared to conventional acoustic liners 
		by simply altering the shape of the orifice or its orientation, in the case of a non-circular orifice. 
		Complementary acoustic simulations demonstrate that this reduced drag may be achieved while maintaining the same noise reduction properties over a wide range of frequencies.
	\end{abstract}
	
	%\section*{Nomenclature}

	{\renewcommand\arraystretch{1.0}
		\noindent\begin{longtable*}{@{}l @{\quad=\quad} l@{}}
			$\omega$  & angular frequency \\
			$A$  & cross-sectional orifice area \\
			$V_c$  & cavity volume \\
			$V_o$  & orifices' volume \\
			$h$  & cavity depth \\
			$t$  & orifice thickness \\
			$c$  & speed of sound \\
			$f$  & frequency of sound wave \\
			$f_r$  & liner resonance frequency \\
			$C_f$  & friction coefficient \\
			$d$  & orifice diameter \\
			$a$  & major axis of ellipse \\
			$b$  & minor axis of ellipse \\
			$h$  & cavity depth \\
			$L_\chi$   & domain length in $\chi$ direction \\
			$M_\infty$ & Freestream Mach number \\
			$M_b$ & Bulk Mach number \\
			$N_\chi$   & grid points in $\chi$ direction \\
			$Re$ & Reynolds number \\
			$u_b$  & bulk velocity \\
			$u_{\text{cl}}$  & centerline velocity \\
			$u_\tau$ & friction velocity \\
			$\alpha$ & Forchheimer permeability \\
			$K$ & Darcy permeability \\	
			$L_\chi$ & Domain size in $\chi$ direction \\
			$\ell_T$ & virtual origin \\
			$\delta$ &    channel half width \\
			$\delta_v$& viscous length scale \\
			$\Delta U^+$  & Hama roughness function \\
			$\lambda_c$ & cavity cross-section length/width \\
			$\nu$ & kinematic viscosity \\
			$\rho$ & density \\
			$\sigma$  & facesheet porosity \\
			$\tau_{ij}$  & Reynolds stresses \\
			$\tau_w$   & wall shear stress \\
	\end{longtable*}}

	\section{Introduction}
	\lettrine{A}{ircraft} engines are the primary source of noise during take-off and landing. 
	To reduce noise, engine nacelles are equipped with noise control devices called acoustic liners.
	Acoustic liners are essentially an array of Helmholtz resonators, which consist of a large cavity with a small opening in the top plate.
	Acoustic liners are simple devices, and their working mechanism is based on the idea
	of dissipating noise by tuning their resonance
	frequency with the dominant frequency of the engine fan.
	They have the potential to attenuate noise by as much as 8--10dB and are thus,
	required on all aircraft jet engines.
	
	Acoustic liners have been optimised primarily from an acoustic perspective~\citep{sivian_acoustic_1935,ingard_acoustic_1950, damiano_2}.
	From an aerodynamic perspective, however, they behave like surface roughness and contribute to an increase in aerodynamic drag.
	Our recent study~\citep{shahzad_turbulence_2023} showed that acoustic liners are responsible for about a 70\% drag increase per plane area
	compared to a hydraulically smooth wall at typical operating conditions.
	This aerodynamic penalty has been accepted as a necessary compromise. 
	
	The effect of acoustic liners on the background turbulent flow has only been studied recently, both using 
	experiments \citep{howerton_acoustic_2015,howerton_conventional_2017, gustavsson_correction_2019,dacome2023innerscaled}
	and numerical simulations \citep{scalo_compressible_2015, zhang_numerical_2016}.
	Experimental studies are often characterized by large uncertainties in drag measurements, and numerical simulations of such surfaces are computationally expensive.
	Therefore, most studies have used simplifying assumptions to make the problem more approachable with numerical simulations. 
	Several authors simulated the flow over a single acoustic liner cavity \citep{avallone_lattice-boltzmann_2019, zhang_numerical_2011, zhang_numerical_2016}
	or replaced the acoustic liner with an equivalent boundary condition~\citep{shur_unsteady_2020,shur_unsteady_2021, sebastian_numerical_2019}.
	As a result, estimates of the added drag reported in the literature have a massive spread, ranging from 3\% up to 500\%, depending
	on the numerical or experimental technique that was used~\citep{shahzad_turbulence_2023}.
	
	Our group recently performed the first Direct Numerical Simulations (DNS) of fully resolved acoustic liner geometries~\citep{shahzad_turbulence_2023}
	and we related the added drag to the wall-normal velocity fluctuations and the wall-normal Forchheimer permeability of the plate. 
	Building on these findings, we showed that it is possible to predict the added drag of acoustic liners in operating conditions~\citep{shahzad_permeability_2022,shahzad_turbulence_2023}.
	Hence, we are confident that better aerodynamic performance can be achieved by finding plate geometries that reduce the wall-normal velocity fluctuations induced by the orifices.
	
	\citet{howerton_acoustic_2016} studied different orifice configurations, changing the orifice shape and orientation, including
	rectangular orifices either parallel or perpendicular to the flow, and found that 
	the perpendicular slot orifice performed better compared to the baseline circular orifices.
	The parallel slot orifice, on the other hand, despite having the same dimensions as the perpendicular slot orifice, had the highest 
	added drag (at a freestream Mach number $M_\infty = 0.3$).
	Furthermore, \citet{howerton_acoustic_2016} noted that, despite changes in the orifice shape, the acoustic performance was largely unchanged,
	meaning that there is potential for reducing the added drag, without hampering the acoustic attenuation properties.
	
	To the best of our knowledge, acoustic liners with varying orifice configurations have been studied only experimentally ~\citep{gaeta_2001,howerton_acoustic_2016}
	and the physical rationale behind their choice of orifice configuration was missing.
	The objective of the current study is to show that it is possible to reduce this aerodynamic penalty while maintaining or improving acoustic noise attenuation.
	In this work, we use our recent findings~\citep{shahzad_turbulence_2023} on scaling laws for the added drag
	to devise optimized acoustic liner geometries that reduce the added drag while having the same acoustic performance as
	the baseline configuration.
	
	\section{Methodology}

	\begin{figure}
		\begin{center}
			\includegraphics{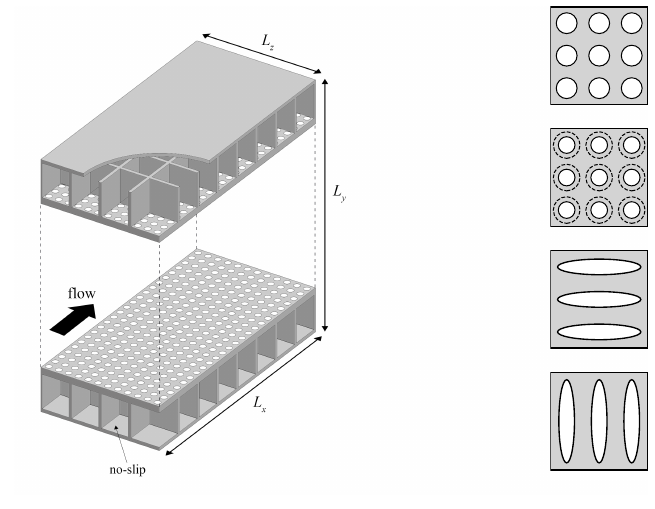}
		\end{center}
		\caption{Sketch of the computational domain.}
		\label{fig:schematic}
	\end{figure}
	
	\begin{table}
		\centering
		\resizebox{0.95\linewidth}{!}{
			\begin{tabular}{l|cccccccccccccc}
				& $Re_b$ & $Re_\tau$ & $d_x^+$ & $d_z^+$ & $K/t^2 \times 10^3$ & $1/(\alpha t)$ & $\Delta x^+$ & $\Delta y^+_\text{min}$ & $\Delta z^+$ & $N_x$ & $N_y$ & $N_z$  & $\Delta U^+$ & $C_f \times 10^3$ \\ \hline
				$S$           & 18536   & 506.1     & 0       & 0      & -      & -            & 5.1          & 0.80     & 5.1          & 300   & 350   & 150   & -      &  4.58    \\
				$L$-$S$       & 16528   & 505.3     & 40.4    & 40.4   & 6.33   & 0.127        & 1.5          & 0.80     & 1.5          & 1000  & 500   & 500   & 1.90   &  5.54    \\
				$L$-$T$       & 16984   & 515.8     & 33.0    & 33.0   & 17.0   & 0.156        & 1.3          & 0.81     & 1.3          & 1200  & 600   & 500   & 1.45   &  5.28    \\
				$L$-$E_{x-3}$ & 13124   & 492.8     & 29.6    & 157.7  & 7.32   & 0.0869       & 5.9          & 0.79     & 1.1          & 250   & 500   & 700   & 4.87   &  7.78    \\
				$L$-$E_{x-6}$ & 15602   & 501.5     & 15.0    & 160.5  & 2.23   & 0.0401       & 6.0          & 0.80     & 0.6          & 250   & 500   & 1300  & 2.33   &  5.80    \\
				$L$-$E_{x-9}$ & 17300   & 515.5     & 10.3    & 165.0  & 1.07   & 0.0217       & 6.2          & 0.82     & 0.4          & 250   & 500   & 2000  & 0.85   &  4.97    \\
				$L$-$E_{z-3}$ & 16104   & 510.0     & 163.2   & 30.6   & 7.32   & 0.0869       & 1.1          & 0.82     & 6.1          & 1400  & 500   & 125   & 2.17   &  5.70    \\
				$L$-$E_{z-6}$ & 16042   & 506.8     & 162.1   & 15.2   & 2.23   & 0.0401       & 0.6          & 0.81     & 6.1          & 2600  & 500   & 125   & 2.29   &  5.78    \\
				$L$-$E_{z-9}$ & 16670   & 515.6     & 165.0   & 10.3   & 1.07   & 0.0217       & 0.4          & 0.82     & 6.2          & 4000  & 500   & 125   & 1.90   &  5.54    \\ \hline
			\end{tabular}
		}
		\caption{DNS dataset comprising smooth ($S$) and liner ($L$-$C$) cases. Simulations are performed at bulk Reynolds number $Re_b = 2u_b\delta/\nu$. $d_x$ and $d_z$ are the lengths of the streamwise and spanwise axes of the orifices. $\Delta x^+$, $\Delta y^+_{\text{min}}$ and $\Delta z^+$ are the viscous-scaled streamwise, minimum wall-normal and spanwise mesh spacing.}
		\label{tab:cases}
	\end{table}

	\begin{figure}
		\begin{center}
			\includegraphics{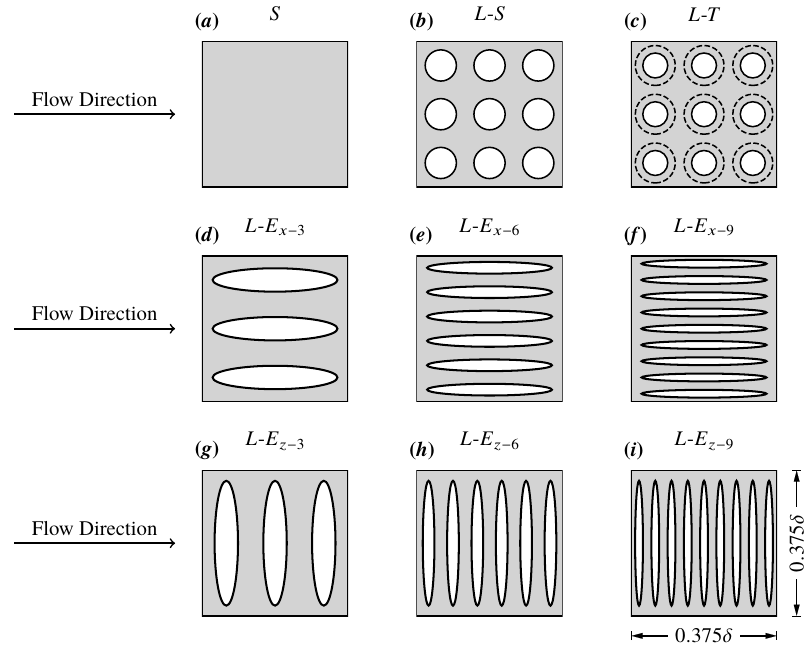}
		\end{center}
		\caption{Orifice configurations considered. All liner cases, except for case $L$-$T$ have a top surface open area ratio of $\sigma = 0.322$.}
			%All liner cases are such that the total volume of the perforations is equal to $V_o=9\pi d_s^2t/4$, where $d_s$ is the diameter of the baseline orifice configurations and $t=d_s$ is the thickness of the facesheet that is constant for all cases.}
		\label{fig:geom}
	\end{figure}
	
	\subsection{Test Setup}
	
	We solve the compressible Navier--Stokes equations for a calorically perfect gas using the solver STREAmS~\citep{bernardini_streams_2021,bernardini_streams_2023} 
	in a plane channel flow configuration.
	The computational domain is a rectangular box of size $L_x \times L_y \times L_z = 3\delta \times 2(\delta+h) \times 1.5\delta$, 
	where $\delta$ is the channel half-width, $h$ is the cavity depth, and $x,y,z$ denote the streamwise, wall-normal and spanwise directions, respectively.
	This domain size is chosen based on previous studies on wall roughness and acoustic liners ~\citep{shahzad_turbulence_2023,chung2015fast,macdonald_17,di_giorgio_relationship_2020,yang_stroh_chung_forooghi_2022}.
	We perform DNS that fully resolves all turbulent length and time scales using a sixth-order spatial discretization and a third-order, three-stage Runge--Kutta scheme for time marching~\citep{spalart_spectral_1991}.
	We carry out all simulations at friction Reynolds numbers $\Rey_\tau = \delta/\delta_v \approx 500$, 
	where $\delta_v$ is the viscous length scale.
	Liner flow cases are complemented by a smooth-wall simulation at approximately the same friction Reynolds number and the same domain size.
	The simulations are performed at bulk Mach number $M_b = u_b/c_w = 0.3$, where $u_b$ is the bulk flow velocity and $c_w$ is the speed of sound at the wall.
	A uniform body force $\Pi$ is added in the streamwise direction and adjusted each time step to maintain a constant mass flow rate in the channel core.
	Conservation of mean-momentum implies that the body force is related to the total drag experienced by the solid wall as $\Pi = \tau_w/\delta$, 
		where $\tau_w$ is the drag per plane area.
	The acoustic liner geometry is handled using a ghost-point forcing immersed boundary method~\citep{vanna_sharp-interface_2020}.
	The immersed boundary method has been previously verified and validated and has been used to study a wide range of geometries~\citep{modesti_22,shahzad_turbulence_2023}.
	Both walls are covered by an array of $8 \times 4$ acoustic liner cavities, as illustrated in Fig. \ref{fig:schematic}.
	Each cavity has a square cross-section with a side length $\lambda_c = 0.335 \delta$ and depth $h = 0.5 \delta$.
	The cavity walls have a thickness of $0.02 \delta$. 
	We use uniform mesh spacing in the streamwise and spanwise directions. 
	In the wall-normal direction, the mesh is clustered towards the wall and coarsened towards the backplate and the channel centre. 
	A minimum of 26 mesh points are used to resolve the orifice diameter (or minor axis in the case of an ellipse).
	This resolution is well within the viscous spacing typically accepted in DNS, and it has been previously validated~\citep{shahzad_turbulence_2023}.
	
	We wish to study the influence of orifice geometry on the aerodynamic and acoustic performance of a liner and compare it to the baseline configuration
	with circular straight holes that are common in most applications.
	For the baseline reference case, see Fig. 2 \textbf{(\textit{b})} we consider the geometry studied by \citet{shahzad_turbulence_2023}
	with porosity $\sigma=0.322$, viscous-scaled diameter $d^+=d/\delta_v \approx 40$ and a thickness to diameter ratio $t/d = 1$, 
	where the orifice diameter for the baseline case is $d = 0.08 \delta$. 
	We chose the baseline liner geometry to be representative of acoustic liners in operating conditions, while also keeping a reasonable computational 
		cost to perform a parametric study using DNS. The orifice size and facesheet thickness are in the range found on engine-mounted acoustic liners
		when scaled with the local boundary layer thickness in landing conditions, as discussed in~\citet{shahzad_turbulence_2023}.
		The cavity depth we use is smaller than in real applications, however, previous studies have reported that its influence on the added drag is negligible~\citep{howerton_acoustic_2015}.
		The porosity is higher than typically found on modern aircraft engines.
		However, in our recent study \citep{shahzad_turbulence_2023} we demonstrate
		that the relevant nondimensional parameter for quantifying the added drag is the viscous-scaled Forchheimer permeability, which is thus a more relevant
		nondimensional quantity than the porosity.

	To compare the result of liner simulations with the smooth wall data, we take into account the effect of the virtual origin $\ell_T$,
	namely, the distance below the plate at which the flow perceives the equivalent smooth wall.
	More details on how the virtual origin is estimated are provided in our previous work~\citep{shahzad_turbulence_2023}.
	In the following, quantities that are non-dimensionalised by $\delta_v$ and $u_{\tau}$ are denoted by the `$+$' superscript.
	The overline symbol $\overline{f}$ is used to indicate Reynolds averages, whereas the tilde $\widetilde{f}=\overline{\rho f}/\overline{\rho}$ 
	indicates Favre averaging, and the double prime symbol indicates fluctuations thereof $f''=f-\widetilde{f}$,
	where $\rho$ is the density.
	As an example, with this notation, the ensemble-averaged Reynolds stress tensor is $\tau_{ij}=\overline{\rho}\widetilde{u_i''u_j''}$,
	where $u_i = \{u_1,u_2,u_3\} = \{u,v,w\}$ are the streamwise, wall-normal and spanwise velocity, respectively.
	All simulations are advanced forward in time until they reach a statistically stationary state, after which statistics are collected for at least $T_{\text{av}} u_\tau/\delta \approx 16$,
		where $T_{\text{av}}$ is the averaging interval.

	\subsection{Novel Configurations}
	
	The facesheet thickness and the cavity dimensions remain unchanged for all cases considered.
	The novel configurations proposed only differ in the orifice shape, size, and orientation.
	In an attempt to reduce the aerodynamic drag induced by acoustic liners, we pursue two ideas:
	
	\begin{itemize}
		\item We aim at increasing the wall-normal Forchheimer permeability of the plate $\alpha_y$.
		Although $\alpha_y$ has a complex dependency on the plate geometry, a first-order approximation is $\alpha_y\sim1/(\sigma^2t)$~\citep{shahzad_permeability_2022}, 
		and therefore a reduction of the plate permeability will result in lower drag.  
		\item We take inspiration from riblets, namely streamwise-aligned surface grooves that are able to reduce friction drag~\citep{modesti_21,endrikat_21},
		and argue that the same surface anisotropy in the streamwise direction might be beneficial for acoustic liners.
	\end{itemize}
	
	Following these two hypotheses, we propose liner geometries that should be more efficient from an aerodynamic perspective and possibly retain the acoustic properties of the liner.
	Therefore, we change the orifice geometry while keeping constant the resonance frequency of the resonators $\omega_r = c_w \sqrt{ {A}/{t V_c}}$,
	where $A$ is the plane area of the orifice, and $V_c$ is the volume of the cavity.
	In this way, we aim to optimise the aerodynamic performance without compromising the acoustic properties.
	
	Based on the first idea, we propose a `tapered-hole' configuration where the orifice has a smaller diameter at the top
	of the facesheet and a larger diameter at the bottom of the facesheet, such that the total volume of the orifice is constant
	and the resonance frequency of the liner, disregarding entry and exit effects, does not change.
	Entry and exit effects are included using empirical correction factors when determining the resonance frequency of the liner as they change the effective mass of air that oscillates in the orifice.
		For a canonical orifice shape, the correction is known and well-documented \citep{rayleigh1871v, ingard_theory_1953, guess1975calculation, zhang_numerical_2016}. Since entry/exit effects are accounted for by empirical corrections, our hypothesis can only be verified a posteriori
		through acoustic simulations at different frequencies.
	For the considered case (case $L$-$T$) the orifice diameter increases continuously from $d=0.064\delta$ at the top of the facesheet to $d=0.1024\delta$ at the bottom of the facesheet.
	
	Based on the second idea, we propose elliptical orifices, that have the same porosity as the baseline liner, i.e. $\sigma = 0.322$.
	The major axis of the ellipse is fixed at $a=0.32\delta$ and
	the minor axis is calculated by assuming a constant porosity and depends on the number of orifices per cavity.
	For example, in the case of 9 elliptical orifices per cavity, see Fig. \ref{fig:geom} \textbf{(\textit{f})} or \textbf{(\textit{i})}, the minor axis is $b=0.02\delta$,
	and in the case of 3 elliptical orifices per cavity, see Fig. \ref{fig:geom} \textbf{(\textit{d})} or \textbf{(\textit{g})}, the minor is $b=0.06\delta$, such that the porosity is always $\sigma = 0.322$.
	
	In addition, the influence of the elliptical orifices' orientation is also studied. While we expect that streamwise-aligned ellipses
	should minimize the added drag, previous experiments~\citep{howerton_acoustic_2016} have shown that rectangular slots perpendicular to the flow have lower drag compared to the canonical configuration. Therefore, we consider elliptical orifices with the major axis aligned with
	the streamwise or spanwise direction. 
	
	The geometries considered are shown in Fig. \ref{fig:geom}, and
	details of all flow cases are reported in Table~\ref{tab:cases}. 
	The naming of the flow cases is as follows: we use the letter $S$
	for the smooth wall case and $L$-$C$ for the liner cases, where $C=\{S,T,E_{x-\chi},E_{z-\chi}\}$ refers to the specific liner flow case.
	In particular, $S$, $T$, $E_{x-\chi}$ and $E_{z-\chi}$ are the baseline liner with straight orifice, tapered-orifice liner, streamwise-oriented ellipses and spanwise-oriented ellipses, respectively, and $\chi$ is the number of ellipses per cavity.

	\section{Results}
	
	\subsection{Aerodynamic Drag}

	\begin{figure}
		\centering
		\begin{tabular}{rr}
			\includegraphics{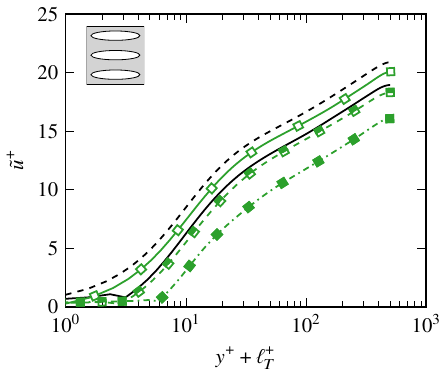}
			\put(-210,185){\textbf{(\textit{a})}} &
			\includegraphics{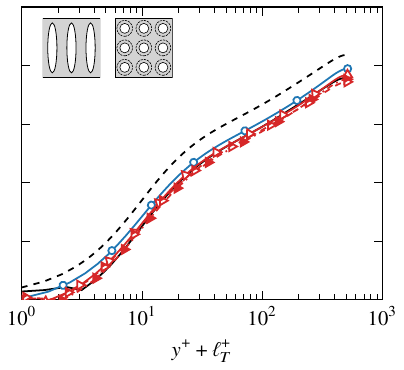} 
			\put(-200,185){\textbf{(\textit{b})}} \\
		\end{tabular}
		\caption{Mean streamwise velocity for streamwise-oriented ellipses as a function of $y^+$. Novel geometries are compared to the smooth wall flow case (dashed black line without symbols) and the baseline liner flow case (solid black line without symbols). The novel liner flow cases have the following line style: $L$-$T$ (circles), $L$-$E_{x-\chi}$ (squares) and $L$-$E_{z-\chi}$ (triangles). Different line types with symbols indicate the number of orifices per cavity: $\chi=3$ (dash-dotted line with filled symbols), $\chi = 6$ (dashed line with half-filled symbols) and $\chi=9$ (solid line with empty symbols).}
		\label{fig:mean_vel}
	\end{figure}

	Figure~\ref{fig:mean_vel} shows the mean streamwise velocity profiles for all flow cases.
	We see that the mean velocity profile over acoustic liners is shifted downward as compared to the smooth wall.
		This downward shift is referred to as the $\Delta U^+$, and it is directly related to the added drag through 
		the exact relation~\citep{shahzad_turbulence_2023},
		\begin{equation}
		\Delta D = 1-\frac{C_f}{C_{f,s}} = 1 - \left(\frac{1}{1-\frac{\Delta U^+}{\sqrt{2/C_{f,s}}}}\right)^2,	
		\label{eq:drag}
		\end{equation}
		where $C_f=2\Pi\delta/(\rho U_{\text{c}}^2)$ is the friction coefficient, $U_{\text{c}}$ is the mean streamwise velocity at the channel centerline,
		and the subscript `s' denotes smooth wall values.
		The advantage of using $\Delta U^+$, instead of the relative variation of the friction coefficient, is that $\Delta U^+$ is independent 
		from the Reynolds number, thus it can be used as a drag measure in laboratory experiments or simulations which necessarily have a lower Reynolds number
		than acoustic liners in operating conditions. 
		The effect of Reynolds number on $\Delta D$ is embedded in $C_{f,s}$, which can be easily estimated using smooth-wall formulas,
		allowing one to use low Reynolds number data to estimate the drag variation in operating conditions, as discussed by~\citet{shahzad_turbulence_2023}.
		A positive $\Delta U^+$ is correlated with a drag increase and represents a downward shift of the velocity profile with respect to the smooth wall.
	In Fig.~\ref{fig:mean_vel} we note that all liner cases show a downward shift compared to the smooth wall ($\Delta U^+$),
	indicating that all cases increase drag.
	However, several liner geometries exhibit a lower $\Delta U^+$ compared to the baseline case,
	demonstrating that modifying the orifice shape can result in a lower added drag.
	For convenience, the value of the Hama roughness function $\Delta U^+$ and the friction coefficient are also reported in Table~\ref{tab:cases}.
	
	Some elliptical orifice configurations show potential for decreasing the added drag.
	However, this depends both on the ellipse dimensions and orientation.
	For streamwise-oriented slots, the drag variation strongly depends on the spanwise size of the orifices. The narrow orifices of flow case $L$-$E_{x-9}$ lead to a substantially lower drag than the baseline case,
	whereas the wider slots of case $L$-$E_{x-3}$ result in a massive drag increase. It is interesting to note that
	the elliptical slots of $L$-$E_{x-9}$ have a spanwise width of size $d_z^+ \approx 10$, which is similar to the spacing of drag-reducing riblets~\citep{modesti_21}.
	For spanwise-oriented slots, $\Delta U^+$ is less sensitive to the slots size, and we find the same or a marginally lower drag than the baseline liner.
	These findings confirm the experiments of \citet{howerton_acoustic_2016} who observed lower drag for spanwise-oriented rectangular slots.
	They found that streamwise-oriented rectangular slots increased drag compared to the baseline case, however, 
	they did not investigate the effect of slot size, which might have been too large in viscous units to observe the `riblets-like' effect we report in this study.
	The tapered orifice also decreases drag compared to the baseline case.
	In this case, the improved performance can be traced back to the reduced superficial porosity experienced by the flow.

	\begin{figure}[p]
		\begin{center}
			\includegraphics{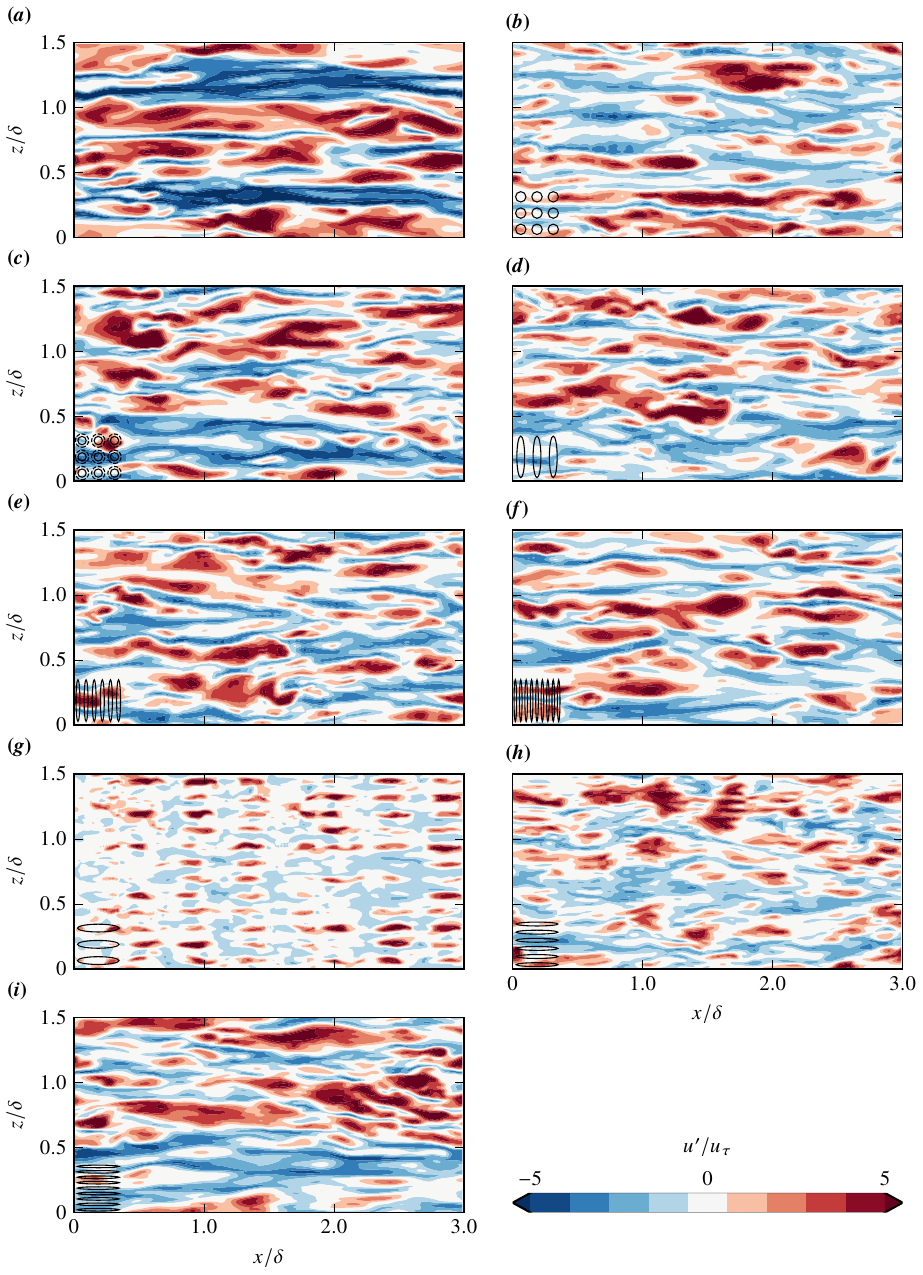}
		\end{center}
		\caption{Streamwise velocity fluctuations on a wall-parallel plane at $y^+ + \ell_T^+ = 8$ for cases $S$ (\textit{a}), $L$-$S$ (\textit{b}), $L$-$T$ (\textit{c}), $L$-$E_{z-3}$ (\textit{d}), $L$-$E_{z-6}$ (\textit{e}), $L$-$E_{z-9}$ (\textit{f}), $L$-$E_{x-3}$ (\textit{g}), $L$-$E_{x-6}$ (\textit{h}), $L$-$E_{x-9}$ (\textit{i}).}
		\label{fig:u_inst}
	\end{figure}

	\begin{figure}[p]
		\begin{center}
			\includegraphics{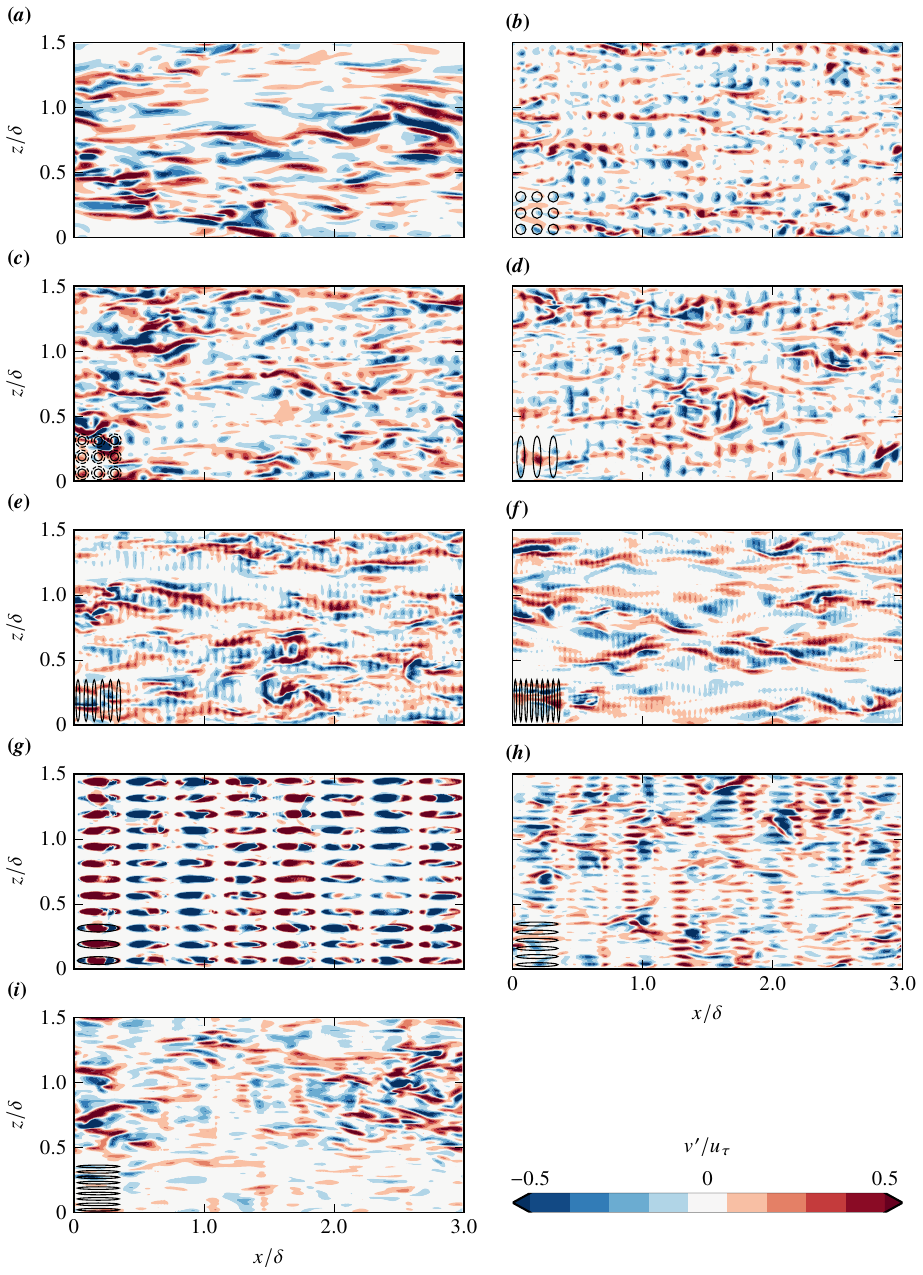}
		\end{center}
		\caption{Wall-normal velocity fluctuations on a wall-parallel plane at $y^+ + \ell_T^+ = 8$ for cases $S$ (\textit{a}), $L$-$S$ (\textit{b}), $L$-$T$ (\textit{c}), $L$-$E_{z-3}$ (\textit{d}), $L$-$E_{z-6}$ (\textit{e}), $L$-$E_{z-9}$ (\textit{f}), $L$-$E_{x-3}$ (\textit{g}), $L$-$E_{x-6}$ (\textit{h}), $L$-$E_{x-9}$ (\textit{i}).}
		\label{fig:v_inst}
	\end{figure}
	
	\begin{figure}[t]
		\centering
		\begin{tabular}{rr}
			\includegraphics{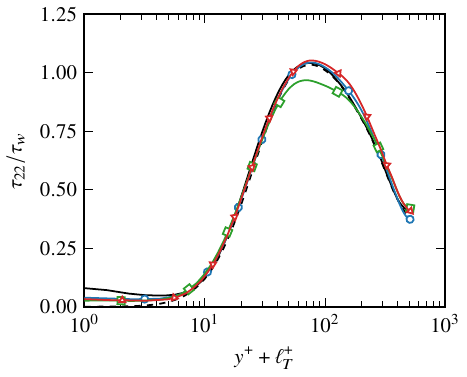}
			\put(-210,185){\textbf{(\textit{f})}} &
			\includegraphics{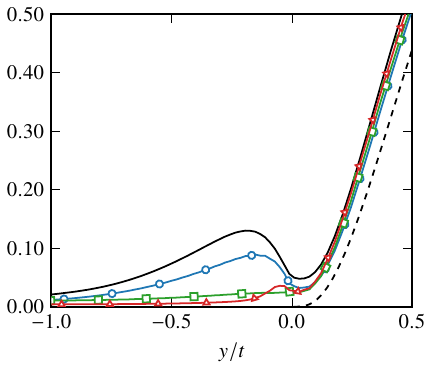} 
			\put(-210,185){\textbf{(\textit{b})}} \\
		\end{tabular}
		\caption{Wall normal Reynolds stress $\tau_{22}$, as a function of $y^+$ (\textit{a}) and as a function of $y/t$ (\textit{b}). Line types and symbols are as in Fig. \ref{fig:mean_vel}.}
		\label{fig:mean_vv}
	\end{figure}
	
	\begin{figure}[t]
		\centering
		\begin{tabular}{rr}
			\includegraphics{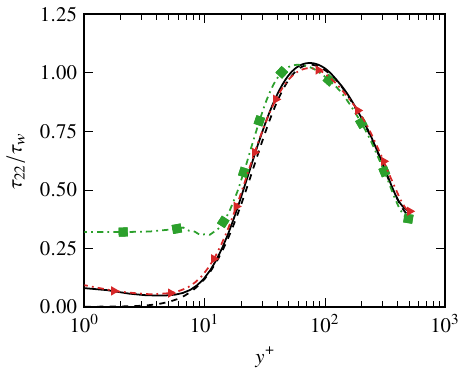}
			\put(-210,185){\textbf{(\textit{a})}} &
			\includegraphics{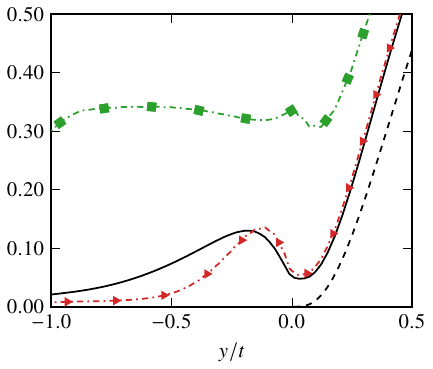} 
			\put(-210,185){\textbf{(\textit{b})}} \\
		\end{tabular}
		\caption{Wall normal Reynolds stress $\tau_{22}$, as a function of $y^+$ (\textit{a}) and as a function of $y/t$ (\textit{b}). Line types and symbols are as in Fig. \ref{fig:mean_vel}.}
		\label{fig:mean_vv_2}
	\end{figure}

	In a previous study, we related the added drag induced by acoustic liners to the wall-normal velocity fluctuations~\citep{shahzad_turbulence_2023};
	thus, we inspect instantaneous velocity realizations in wall parallel planes at $y^+ + \ell_T^+ = 8$ in Fig.~\ref{fig:u_inst} and \ref{fig:v_inst}. 
	For all liner cases, we note that near-wall streaks are disrupted and much shorter than on a smooth wall, see Fig.~\ref{fig:u_inst}.
	The break-up of the near-wall streaks is particularly evident for case $L$-$E_{x-3}$ (Fig.~\ref{fig:u_inst} \textbf{(\textit{g})}), where the near-wall flow deviates substantially from the 
	typical organization found for the smooth wall (Fig.~\ref{fig:u_inst} \textbf{(\textit{a})}).
	
	For the baseline liner geometry, wall-normal velocity fluctuations are concentrated, primarily, around the orifice location, 
	and the positions of the orifices are clearly visible in the contours of the wall-normal velocity, see Fig.~\ref{fig:v_inst}~\textbf{(\textit{b})}. 
	This effect is more evident for cases $L$-$E_{x-3}$ and $L$-$E_{x-6}$, 
	where the wall-normal velocity fluctuations in the near wall region originate primarily from the orifices.
	The higher wall-normal velocity fluctuations for flow case $L$-$E_{x-3}$ are correlated with the significantly higher drag of this case.
	
	Furthermore, we find that liner cases with lower drag present lower wall-normal velocity fluctuations and a more smooth-wall-like organization of the near-wall flow.
	These qualitative observations suggest that the correlation formulated by \citet{shahzad_turbulence_2023} that relates the added drag and the intensity of the wall-normal velocity fluctuations holds
	for all facesheet geometries under scrutiny.

	For a more quantitative analysis,
	we further show the wall-normal Reynolds stress components $\tau_{22} = \overline{\rho} \widetilde{v''v''}$ in Fig.~\ref{fig:mean_vv} and ~\ref{fig:mean_vv_2}. 
	It is clear that the wall-normal velocity fluctuations are not zero, irrespective of the case considered, above and below the facesheet.
	We find that flow cases that exhibit lower drag than the baseline liner show lower wall-normal velocity fluctuations in the near-wall region
	and below the facesheet, Fig.~\ref{fig:mean_vv}.
	The converse is true for cases that increase the added drag, Fig.~\ref{fig:mean_vv_2}.

	\subsection{Acoustic Attenuation}

	\begin{figure}
		\begin{center}
			\includegraphics{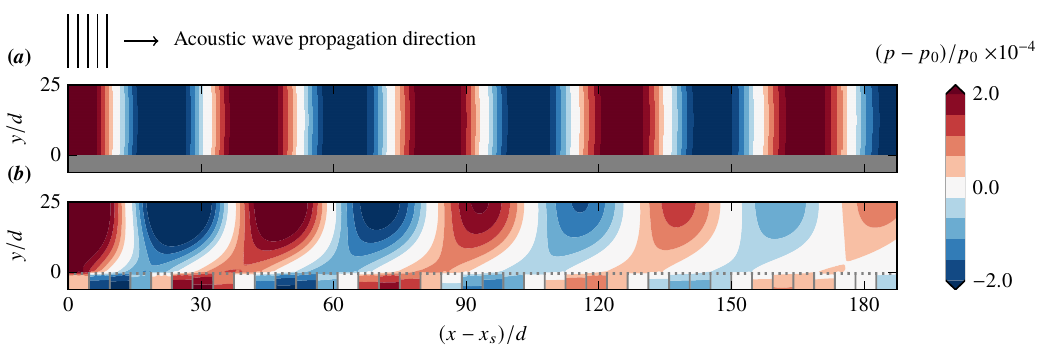} 
		\end{center}
		\caption{Instantaneous pressure fluctuations over the smooth wall (\textit{a}) and the baseline liner (\textit{b}). $p_0$ is the reference thermodynamic pressure. 
		 $x_s$ is the streamwise location where the liners start. The frequency of the sound waves is $f/f_r = 1$.}
		\label{fig:spl_inst}
	\end{figure}
	
	\begin{figure}
		\begin{center}
			\includegraphics{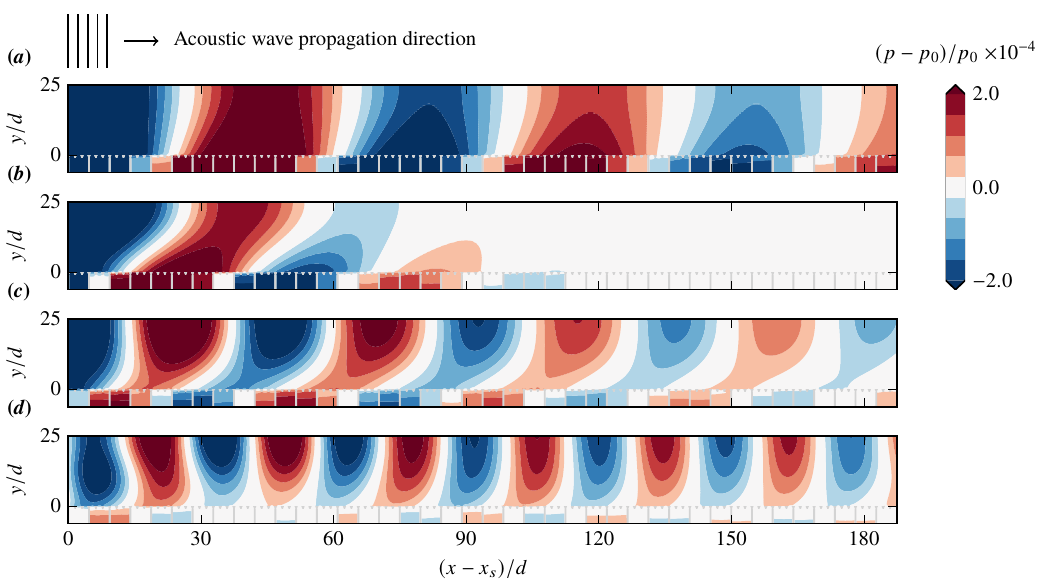}
		\end{center}
		\caption{Instantaneous pressure fluctuations for the tapered orifice configuration for various acoustic wave frequencies $f/f_r$= $0.5$ (\textit{a}), $0.75$ (\textit{b}), $1.0$ (\textit{c}) and $1.5$ (\textit{d}).}
		\label{fig:spl_inst2}
	\end{figure}
	
	\begin{figure}
		\centering
		\begin{tabular}{lr}
			\includegraphics{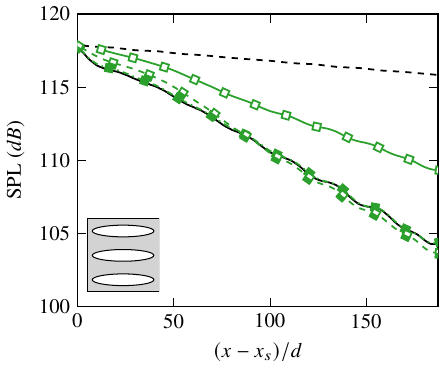}
			\put(-210,185){\textbf{(\textit{a})}} &
			\includegraphics{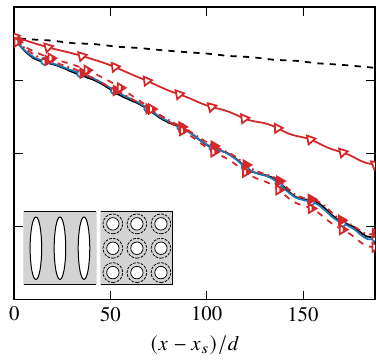}
			\put(-190,185){\textbf{(\textit{b})}} \\
		\end{tabular}
		\caption{Sound pressure level as a function of the streamwise distance from the start of the liners. Line types and symbols are as in Fig. \ref{fig:mean_vel}.}% $L$-$T$ (circles), $L$-$E_{x-\chi}$ (squares) and $L$-$E_{z-\chi}$ (triangles). Different line types with symbols indicate the number of orifices per cavity: $\chi=3$ (dash-dotted line with filled symbols), $\chi = 6$ (dashed line with half-filled symbols) and $\chi=9$ (solid line with empty symbols).}
		\label{fig:spl}
	\end{figure}
	
	\begin{figure}
		\centering
		\begin{tabular}{lr}
			\includegraphics{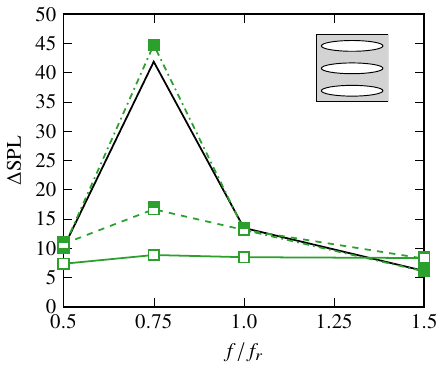} 
			\put(-210,185){\textbf{(\textit{a})}} &
			\includegraphics{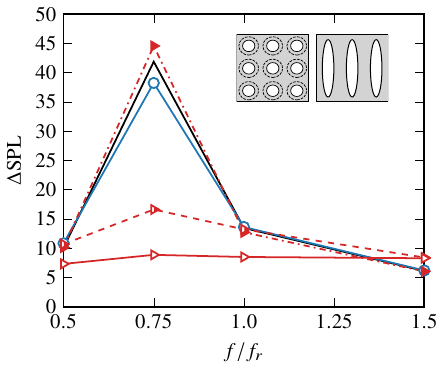}
			\put(-190,185){\textbf{(\textit{b})}} \\
		\end{tabular}
		\caption{Sound pressure level as a function of the forcing frequency. Line types and symbols are as in Fig. \ref{fig:mean_vel}.}% $L$-$T$ (circles), $L$-$E_{x-\chi}$ (squares) and $L$-$E_{z-\chi}$ (triangles). Different line types with symbols indicate the number of orifices per cavity: $\chi=3$ (dash-dotted line with filled symbols), $\chi = 6$ (dashed line with half-filled symbols) and $\chi=9$ (solid line with empty symbols).}
		\label{fig:spl2}
	\end{figure}

	We have studied the aerodynamic drag of different orifice configurations and have identified several geometries
	that induce lower drag compared to the baseline liner with circular orifices.
	As acoustic liners are used to reduce engine noise, it is imperative to also test their acoustic performance.
	In order to do so, we study the noise attenuation properties of these novel configurations in the absence of grazing flow.
	A complete picture of the noise attenuation properties of aerodynamically optimised acoustic liner geometries would require including the influence of grazing flow.
	However, studies of acoustic liners without grazing flow have been extensively used to provide an initial estimation of their acoustic performance \citep{tam_computational_2010, zhang_numerical_2012, schroeder_numerical2021}.
	Simulations in the absence of grazing flow allow us to study the acoustic performance over a range of frequencies for all geometries considered at tractable computational cost.
	
	Simulations are performed in a channel of size $L_x \times L_y \times L_z = 1250 d \times (25d + h) \times 4.69d$, where $d$ is the orifice diameter of the baseline liner and $h=6.25d$ is the depth of the liner.
	The spanwise domain size corresponds to a single cavity, and periodic boundary conditions are applied in the spanwise direction.
	We apply no-slip isothermal boundary conditions on the top and bottom walls and place acoustic liners at the bottom wall of the channel. 
	An array of $40 \times 1$ acoustic liner cavities is placed between $x_s = 500d$ and $x_e = 687.5d$.
	The geometries of the liners are the same as those considered when studying the aerodynamic performance.
	The resonance frequency of the system is expected to be approximately $f_r \approx 0.028c/d$.
		However, the correction due to the pressure field, particularly for the geometries considered in the paper, is difficult to estimate.
		We, therefore, test the performance of the novel geometries over a range of frequencies from $f/f_r = 0.5$ to $f/f_r = 1.5$
	at a Reynolds number $Re_c = 800$ based on the orifice diameter and speed of sound.
	We verified that, while the absolute sound attenuation of these liners changes,
	the relative sound attenuation, compared to the baseline acoustic liner, does not depend on the Reynolds number.
	
	Figure \ref{fig:spl_inst} shows an instantaneous snapshot of the pressure for the smooth wall case and the baseline liner, for the case with $f = f_r$.
	Figure \ref{fig:spl_inst2} also shows how the instantaneous pressure field changes as the forcing frequency is changed.
	As the frequency of the acoustic wave is tuned to the resonant frequency of the liner, we see a significant attenuation of the acoustic wave over the liner. 
	The amplitude of the pressure fluctuations towards the end of the acoustic liner is much lower than over the smooth wall, where the fluid viscosity
	is the only dissipative mechanism.
	
	Figure \ref{fig:spl} compares the SPL evolution of the acoustic wave for all the liner configurations and the smooth wall,
	for the case with $f = f_r$.
	Although there is a clear SPL reduction for all configurations as compared to the smooth wall, some geometries perform better than others
	when evaluated against the baseline liner.
	This is true for the SPL loss over the entire frequency range considered, see Fig. \ref{fig:spl2}.
		Cases $L$-$E_{x-3}$ and $L$-$E_{z-3}$, offer very similar acoustic noise attenuation over the entire frequency range, however, 
		the narrow orifices of cases $L$-$E_{x-6}$, $L$-$E_{z-6}$, $L$-$E_{x-9}$ and $L$-$E_{z-9}$, that helped reduce drag,
		appear to inhibit acoustic noise attenuation.
		Cases $L$-$E_{x-6}$ and $L$-$E_{z-6}$ show similar noise attenuation properties at frequencies other than $f = 0.75f_r$,
		whereas cases $L$-$E_{x-9}$ and $L$-$E_{z-9}$ show significantly lower attenuation throughout the frequency range.
		The tapered hole configuration provides comparable acoustic noise attenuation as compared to the standard acoustic liner.
	Figure \ref{fig:spl2} also shows that the effective resonance frequency is about 25\% lower than the nominal one, that we estimated disregarding entrance and exit effects.
		Notably, the resonant frequency appears to be about the same for all orifice shapes, suggesting that the required correction for entrance and exit effects is similar for all 
		orifice shapes.
	The tapered holes, therefore, improve not only the aerodynamic performance of the liner but also
	perform well acoustically.
	The elliptical orifices, however, depending upon the width of the orifice, may or may not have a detrimental influence on acoustic noise attenuation.

	\section{Conclusion}
	
	We have proposed several novel orifice geometries for acoustic liners, some of which reduce aerodynamic drag without compromising the acoustic performance.
	The idea behind the aerodynamically optimized geometries is based on flow physics, which allows us to restrict the vast parameter space that one could explore.
	The aerodynamic performance of the novel geometries is scrutinised based on Direct Numerical Simulations of fully resolved acoustic liner arrays in a turbulent channel flow.
	We find that tapered circular orifices minimize drag compared to a baseline acoustic liner by reducing the apparent porosity at the surface of the facesheet, whereas
	replacing the circular holes with elliptical slots can lead to a substantially lower drag if the minor axis of the ellipse is sufficiently thin in viscous units.
	We find that despite the very different configurations tested, all optimised geometries work by reducing the interaction of the flow above and below the surface of the facesheet,
	which is confirmed by the reduced wall-normal velocity fluctuations in the proximity of the facesheet.
	
	We also test the acoustic performance of the proposed liner configurations and find that the thin elliptical slots have a substantially reduced acoustic performance compared to 
	a baseline liner; therefore, the improved aerodynamic benefit comes at the cost of lower noise reduction.
	The tapered hole configuration has slightly better noise attenuation properties than the baseline liner while offering substantially lower aerodynamic drag,
	and it is therefore superior both from an aerodynamic and acoustic perspective. In addition, tapered holes are easy to manufacture and
	therefore represent a viable modification to be implemented in existing designs.

	\section*{Acknowledgments}
	We acknowledge EuroHPC for awarding us access to Meluxina, at LuxProvide, Luxembourg.

	{\bf Data availability statement}
	
	DNS data is available at \href{https://doi.org/10.4121/1b59fc2b-d837-4479-836e-810aa5b0db3d}{https://doi.org/10.4121/1b59fc2b-d837-4479-836e-810aa5b0db3d}
	
	\vspace{24pt}

	\bibliography{sample}

\begin{thebibliography}{35}
\newcommand{\enquote}[1]{``#1''}
\providecommand{\natexlab}[1]{#1}
\providecommand{\url}[1]{\texttt{#1}}
\providecommand{\urlprefix}{URL }
\expandafter\ifx\csname urlstyle\endcsname\relax
  \providecommand{\doi}[1]{\discretionary{}{}{}https://doi.org/#1}\else
  \providecommand{\doi}[1]{\discretionary{}{}{}\urlstyle{rm}\url{https://doi.org/#1}}\fi

\bibitem[{Sivian(1935)}]{sivian_acoustic_1935}
Sivian, L.~J., \enquote{Acoustic {Impedance} of {Small} {Orifices},} \emph{J.
  Acoust. Soc}, Vol.~7, No.~2, 1935, pp. 94--101.
\newblock \doi{10.1121/1.1915795}.

\bibitem[{Ingård and Labate(1950)}]{ingard_acoustic_1950}
Ingård, U., and Labate, S., \enquote{Acoustic {Circulation} {Effects} and the
  {Nonlinear} {Impedance} of {Orifices},} \emph{J. Acoust. Soc. Am.}, Vol.~22,
  No.~2, 1950, pp. 211--218.
\newblock \doi{10.1121/1.1906591}.

\bibitem[{Mann et~al.(2013)Mann, Perot, Kim, and Casalino}]{damiano_2}
Mann, A., Perot, F., Kim, M.-S., and Casalino, D., \enquote{Characterization of
  Acoustic Liners Absorption using a Lattice-Boltzmann Method,} \emph{{AIAA}
  paper 2013-2271}, 2013.
\newblock \doi{10.2514/6.2013-2271}.

\bibitem[{Shahzad et~al.(2023)Shahzad, Hickel, and
  Modesti}]{shahzad_turbulence_2023}
Shahzad, H., Hickel, S., and Modesti, D., \enquote{Turbulence and added drag
  over acoustic liners,} \emph{J. Fluid Mech.}, Vol. 965, 2023, p. A10.
\newblock \doi{10.1017/jfm.2023.397}.

\bibitem[{Howerton and Jones(2015)}]{howerton_acoustic_2015}
Howerton, B.~M., and Jones, M.~G., \enquote{Acoustic {liner} {drag}: {a}
  {parametric} {study} of {conventional} {configurations},} \emph{{AIAA} paper
  2015-2230}, 2015.
\newblock \doi{10.2514/6.2015-2230}.

\bibitem[{Howerton and Jones(2017)}]{howerton_conventional_2017}
Howerton, B.~M., and Jones, M.~G., \enquote{A {conventional} {liner}
  {acoustic}/{drag} {interaction} {benchmark} {database},} \emph{{AIAA} paper
  2017-4190}, 2017.
\newblock \doi{10.2514/6.2017-4190}.

\bibitem[{Gustavsson et~al.(2019)Gustavsson, Zhang, Cattafesta, and
  Kreitzman}]{gustavsson_correction_2019}
Gustavsson, J., Zhang, Y., Cattafesta, L.~N., and Kreitzman, J.~R.,
  \enquote{Acoustic {liner} {drag} {measurements},} \emph{{AIAA} paper
  2019-2683}, 2019.
\newblock \doi{10.2514/6.2019-2683.c1}.

\bibitem[{Dacome et~al.(2023)Dacome, Siebols, and
  Baars}]{dacome2023innerscaled}
Dacome, G., Siebols, R., and Baars, W.~J., \enquote{Inner-scaled Helmholtz
  resonators with grazing turbulent boundary layer flow,}
  \emph{arXiv:2308.07776 [physics]}, 2023.
\newblock \doi{10.48550/arXiv.2308.07776}.

\bibitem[{Scalo et~al.(2015)Scalo, Bodart, and Lele}]{scalo_compressible_2015}
Scalo, C., Bodart, J., and Lele, S.~K., \enquote{Compressible turbulent channel
  flow with impedance boundary conditions,} \emph{Phys. Fluids}, Vol.~27,
  No.~3, 2015, p. 035107.
\newblock \doi{10.1063/1.4914099}.

\bibitem[{Zhang and Bodony(2016)}]{zhang_numerical_2016}
Zhang, Q., and Bodony, D.~J., \enquote{Numerical investigation of a honeycomb
  liner grazed by laminar and turbulent boundary layers,} \emph{J. Fluid
  Mech.}, Vol. 792, 2016, pp. 936--980.
\newblock \doi{10.1017/jfm.2016.79}.

\bibitem[{Avallone et~al.(2019)Avallone, Manjunath, Ragni, and
  Casalino}]{avallone_lattice-boltzmann_2019}
Avallone, F., Manjunath, P., Ragni, D., and Casalino, D.,
  \enquote{Lattice-{Boltzmann} {Very} {large} {eddy} {simulation} of a
  {multi}-{orifice} {acoustic} {liner} with {turbulent} {grazing} {flow},}
  \emph{{AIAA} paper 2019-2542}, 2019.
\newblock \doi{10.2514/6.2019-2542}.

\bibitem[{Zhang and Bodony(2011)}]{zhang_numerical_2011}
Zhang, Q., and Bodony, D.~J., \enquote{Numerical {simulation} of
  {two}-{dimensional} {acoustic} {liners} with {high}-{speed} {grazing}
  {flow},} \emph{AIAA Journal}, Vol.~49, No.~2, 2011, pp. 365--382.
\newblock \doi{10.2514/1.J050597}.

\bibitem[{Shur et~al.(2020)Shur, Strelets, Travin, Suzuki, and
  Spalart}]{shur_unsteady_2020}
Shur, M., Strelets, M., Travin, A., Suzuki, T., and Spalart, P.~R.,
  \enquote{Unsteady {simulation} of {sound} {propagation} in {turbulent} {flow}
  {inside} a {lined} {duct} {using} a {broadband} {time}-{domain} {impedance}
  {model},} \emph{{AIAA} paper 2020-2535}, 2020.
\newblock \doi{10.2514/6.2020-2535}.

\bibitem[{Shur et~al.(2021)Shur, Strelets, Travin, Suzuki, and
  Spalart}]{shur_unsteady_2021}
Shur, M., Strelets, M., Travin, A., Suzuki, T., and Spalart, P.,
  \enquote{Unsteady {Simulations} of {Sound} {Propagation} in {Turbulent}
  {Flow} {Inside} a {Lined} {Duct},} \emph{AIAA Journal}, 2021, pp. 1--17.
\newblock \doi{10.2514/1.J060181}.

\bibitem[{Sebastian et~al.(2019)Sebastian, Marx, and
  Fortuné}]{sebastian_numerical_2019}
Sebastian, R., Marx, D., and Fortuné, V., \enquote{Numerical simulation of a
  turbulent channel flow with an acoustic liner,} \emph{J. Sound Vib.}, Vol.
  456, 2019, pp. 306--330.
\newblock \doi{10.1016/j.jsv.2019.05.020}.

\bibitem[{Shahzad et~al.(2022)Shahzad, Hickel, and
  Modesti}]{shahzad_permeability_2022}
Shahzad, H., Hickel, S., and Modesti, D., \enquote{Permeability and turbulence
  over perforated plates,} \emph{Flow. Turbul. Combust.}, Vol. 109, No.~4,
  2022, p. 1241–1254.
\newblock \doi{10.1007/s10494-022-00337-7}.

\bibitem[{Howerton and Jones(2016)}]{howerton_acoustic_2016}
Howerton, B.~M., and Jones, M.~G., \enquote{Acoustic {liner} {drag}:
  {measurements} on {novel} {facesheet} {perforate} {geometries},} \emph{{AIAA}
  paper 2016-2979}, 2016.
\newblock \doi{10.2514/6.2016-2979}.

\bibitem[{R.~Gaeta and Ahuja()}]{gaeta_2001}
R.~Gaeta, J., and Ahuja, K., \enquote{Effect of orifice shape on acoustic
  impedance,} \emph{AIAA paper 2001-02}, ????
\newblock \doi{10.2514/6.2001-662}.

\bibitem[{Bernardini et~al.(2021)Bernardini, Modesti, Salvadore, and
  Pirozzoli}]{bernardini_streams_2021}
Bernardini, M., Modesti, D., Salvadore, F., and Pirozzoli, S.,
  \enquote{{STREAmS}: a high-fidelity accelerated solver for direct numerical
  simulation of compressible turbulent flows,} \emph{Comput. Phys. Commun.},
  Vol. 263, 2021, p. 107906.
\newblock \doi{10.1016/j.cpc.2021.107906}.

\bibitem[{Bernardini et~al.(2023)Bernardini, Modesti, Salvadore,
  Sathyanarayana, {Della Posta}, and Pirozzoli}]{bernardini_streams_2023}
Bernardini, M., Modesti, D., Salvadore, F., Sathyanarayana, S., {Della Posta},
  G., and Pirozzoli, S., \enquote{STREAmS-2.0: Supersonic turbulent accelerated
  Navier-Stokes solver version 2.0,} \emph{Comput. Phys. Commun.}, Vol. 285,
  2023, p. 108644.
\newblock \doi{https://doi.org/10.1016/j.cpc.2022.108644}.

\bibitem[{Chung et~al.(2015)Chung, Chan, MacDonald, Hutchins, and
  Ooi}]{chung2015fast}
Chung, D., Chan, L., MacDonald, M., Hutchins, N., and Ooi, A., \enquote{A fast
  direct numerical simulation method for characterising hydraulic roughness,}
  \emph{J. Fluid Mech.}, Vol. 773, 2015, pp. 418--431.
\newblock \doi{10.1017/jfm.2015.230}.

\bibitem[{MacDonald et~al.(2017)MacDonald, Chung, Hutchins, Chan, Ooi, and
  Garc{\'\i}a-Mayoral}]{macdonald_17}
MacDonald, M., Chung, D., Hutchins, N., Chan, L., Ooi, A., and
  Garc{\'\i}a-Mayoral, R., \enquote{The minimal-span channel for rough-wall
  turbulent flows,} \emph{J. Fluid Mech.}, Vol. 816, 2017, pp. 5--42.
\newblock \doi{10.1017/jfm.2017.69}.

\bibitem[{Di~Giorgio et~al.(2020)Di~Giorgio, Leonardi, Pirozzoli, and
  Orlandi}]{di_giorgio_relationship_2020}
Di~Giorgio, S., Leonardi, S., Pirozzoli, S., and Orlandi, P., \enquote{On the
  relationship between drag and vertical velocity fluctuations in flow over
  riblets and liquid infused surfaces,} \emph{Int. J. Heat Fluid Flow},
  Vol.~86, 2020, p. 108663.
\newblock \doi{10.1016/j.ijheatfluidflow.2020.108663}.

\bibitem[{Yang et~al.(2022)Yang, Stroh, Chung, and
  Forooghi}]{yang_stroh_chung_forooghi_2022}
Yang, J., Stroh, A., Chung, D., and Forooghi, P., \enquote{Direct numerical
  simulation-based characterization of pseudo-random roughness in minimal
  channels,} \emph{J. Fluid Mech.}, Vol. 941, 2022, p. A47.
\newblock \doi{10.1017/jfm.2022.331}.

\bibitem[{Spalart et~al.(1991)Spalart, Moser, and
  Rogers}]{spalart_spectral_1991}
Spalart, P.~R., Moser, R.~D., and Rogers, M.~M., \enquote{Spectral methods for
  the {Navier}-{Stokes} equations with one infinite and two periodic
  directions,} \emph{J. Comput. Phys.}, Vol.~96, No.~2, 1991, pp. 297--324.
\newblock \doi{10.1016/0021-9991(91)90238-G}.

\bibitem[{Vanna et~al.(2020)Vanna, Picano, and
  Benini}]{vanna_sharp-interface_2020}
Vanna, F.~D., Picano, F., and Benini, E., \enquote{A sharp-interface immersed
  boundary method for moving objects in compressible viscous flows,}
  \emph{Comput. Fluids}, Vol. 201, 2020, p. 104415.
\newblock \doi{10.1016/j.compfluid.2019.104415}.

\bibitem[{Modesti et~al.(2022)Modesti, Sathyanarayana, Salvadore, and
  Bernardini}]{modesti_22}
Modesti, D., Sathyanarayana, S., Salvadore, F., and Bernardini, M.,
  \enquote{Direct numerical simulation of supersonic turbulent flows over rough
  surfaces,} \emph{J. Fluid Mech.}, Vol. 942, 2022, p. A44.

\bibitem[{Modesti et~al.(2021)Modesti, Endrikat, Hutchins, and
  Chung}]{modesti_21}
Modesti, D., Endrikat, S., Hutchins, N., and Chung, D., \enquote{Dispersive
  stresses in turbulent flow over riblets,} \emph{J. Fluid Mech.}, Vol. 917,
  2021, p. A55.
\newblock \doi{10.1017/jfm.2021.310}.

\bibitem[{Endrikat et~al.(2021)Endrikat, Modesti, Garc{\'\i}a-Mayoral,
  Hutchins, and Chung}]{endrikat_21}
Endrikat, S., Modesti, D., Garc{\'\i}a-Mayoral, R., Hutchins, N., and Chung,
  D., \enquote{Influence of riblet shapes on the occurrence of
  Kelvin--Helmholtz rollers,} \emph{J. Fluid Mech.}, Vol. 913, 2021, p. A37.
\newblock \doi{10.1017/jfm.2021.2}.

\bibitem[{Rayleigh(1871)}]{rayleigh1871v}
Rayleigh, J.~B., \enquote{On the theory of resonance,} \emph{Philos. Trans. R.
  Soc. A}, Vol. 161, 1871, pp. 77--118.
\newblock \doi{10.1098/rstl.1871.0006}.

\bibitem[{Ingård(1953)}]{ingard_theory_1953}
Ingård, U., \enquote{On the {Theory} and {Design} of {Acoustic} {Resonators},}
  \emph{J. Acoust. Soc. Am.}, Vol.~25, No.~6, 1953, pp. 1037--1061.
\newblock \doi{10.1121/1.1907235}.

\bibitem[{Guess(1975)}]{guess1975calculation}
Guess, A., \enquote{Calculation of perforated plate liner parameters from
  specified acoustic resistance and reactance,} \emph{J. Sound Vib.}, Vol.~40,
  No.~1, 1975, pp. 119--137.
\newblock \doi{10.1016/S0022-460X(75)80234-3}.

\bibitem[{Tam et~al.(2010)Tam, Ju, Jones, Watson, and
  Parrott}]{tam_computational_2010}
Tam, C. K.~W., Ju, H., Jones, M.~G., Watson, W.~R., and Parrott, T.~L.,
  \enquote{A computational and experimental study of resonators in three
  dimensions,} \emph{J. Sound Vib.}, Vol. 329, No.~24, 2010, pp. 5164--5193.
\newblock \doi{10.1016/j.jsv.2010.06.005}.

\bibitem[{Zhang and Bodony(2012)}]{zhang_numerical_2012}
Zhang, Q., and Bodony, D.~J., \enquote{Numerical investigation and modelling of
  acoustically excited flow through a circular orifice backed by a hexagonal
  cavity,} \emph{J. Fluid Mech.}, Vol. 693, 2012, pp. 367--401.
\newblock \doi{10.1017/jfm.2011.537}.

\bibitem[{Schroeder et~al.(2021)Schroeder, Spillere, Bonomo, da~Silva,
  Avallone, and Cordioli}]{schroeder_numerical2021}
Schroeder, L., Spillere, A.~M., Bonomo, L.~A., da~Silva, A.~R., Avallone, F.,
  and Cordioli, J.~A., \enquote{Numerical Investigation of Acoustic Liners
  Experimental Techniques using a Lattice-Boltzmann Solver,} \emph{{AIAA} paper
  2021-2144}, 2021.
\newblock \doi{10.2514/6.2021-2144}.

\end{thebibliography}
	
\end{document}